\documentclass[%
 aip,
 jmp,%
 amsmath,amssymb,
preprint,
]{revtex4-1}

\usepackage{graphicx}
\usepackage{dcolumn}
\usepackage{bm}

\begin{document}

\preprint{AIP/123-QED}

\title[T.Schemme et al.]{Structure and morphology of epitaxially grown Fe$_{3}$O$_{4}$/NiO bilayers on MgO(001)}

\author{T. Schemme}
\email{toschemm@uos.de.}
\affiliation{Fachbereich Physik, Universit\"at Osnabr\"uck, Barbarastr.\,7, 49069 Osnabr\"uck, Germany}
\author{O. Kuschel}
\affiliation{Fachbereich Physik, Universit\"at Osnabr\"uck, Barbarastr.\,7, 49069 Osnabr\"uck, Germany}
\author{F. Bertram}
\affiliation{DESY, Photon Science, Notkestr. 85, 22607 Hamburg, Germany}
\author{K. Kuepper}
\affiliation{Fachbereich Physik, Universit\"at Osnabr\"uck, Barbarastr.\,7, 49069 Osnabr\"uck, Germany}
\author{J. Wollschl\"ager}%
 \email{jwollsch@uos.de.}
\affiliation{Fachbereich Physik, Universit\"at Osnabr\"uck, Barbarastr.\,7, 49069 Osnabr\"uck, Germany}

\date{\today}

\begin{abstract}
Crystalline Fe$_{3}$O$_{4}$/NiO bilayers were grown on MgO(001) substrates using reactive molecular beam epitaxy to investigate their structural properties and their morphology. The film thickness either of the Fe$_{3}$O$_{4}$ film or of the NiO film has been varied to shed light on the relaxation of the bilayer system. The surface properties as studied by x-ray photo electron spectroscopy and low energy electron diffraction show clear evidence of stoichiometric well-ordered film surfaces. Based on the kinematic approach x-ray diffraction experiments were completely analyzed. As a result the NiO films grow pseudomorphic in the investigated thickness range (up to 34\,nm) while the Fe$_{3}$O$_{4}$ films relax continuously up to the thickness of 50\,nm. Although all diffraction data show well developed Laue fringes pointing to oxide films of very homogeneous thickness, the Fe$_{3}$O$_{4}$-NiO interface roughens continuously up to 1\,nm root-mean-square roughness with increasing NiO film thickness while the Fe3O4 surface is very smooth independent on the Fe$_{3}$O$_{4}$ film thickness. Finally, the Fe$_{3}$O$_{4}$-NiO interface spacing is similar to the interlayer spacing of the oxide films while the NiO-MgO interface is expanded.
\end{abstract}

\pacs{Valid PACS appear here}
\maketitle

\section{Introduction}
The modification of magnetic properties of ferro(i)magnetic films (F) by antiferromagnetic films (AF) is of huge physical and technological interest for instance for the development of magnetoresistive (MR) devices like magnetic tunnel junctions (MTJs)\cite{MTJ,MTJsMoodera,OxideSpintronics,SpintronicsApps}. MTJs primarily consist of two ferro(i)magnetic conducting films, which are separated by a non-magnetic insulator. In case both ferro(i)magnetic films are comprised of the same material, it is essential to shift the coercive field of one of the films. The AF/F exchange coupling causing an exchange bias can be utilized to have two different switching fields. Thus, the magnetization of the films can be switched separately and an alternation of the alignment of the magnetization between parallel and antiparallel is possible.\\
Magnetite (Fe$_{3}$O$_{4}$) is a promising material for such applications due its half-metallic\cite{halfmet} character and a high spin polarization at the Fermi level. The ferrimagnetic oxide has a high Curie temperature (860\,K) and a saturation moment of almost 4$\mu_{B}$. It crystallizes in an inverse spinel structure with a cubic lattice constant of 0.83964\,nm at 300\,K. Below the so called Verwey temperature T$_{V}$ = 120\,K occurs a phase transition, whereby magnetite adopts a monoclinic structure, becomes insulting and its susceptibility changes.\\
NiO is an antiferromagnetic ionic insulator with a high thermal stability. It is inert against corrosion and has a N$\rm \acute{e}$el temperature of T$_{N}$ = 523\,K and therefore well suited as exchange bias material  NiO crystallizes in a rock salt structure with a lattice constant of 0.41769\,nm.\\
MgO is isostructural to NiO with a slightly larger lattice constant of 0.42117\,nm. Hence, the lattice mismatch between NiO and MgO is about 0.8\,\% while the misfit between Fe$_{3}$O$_{4}$ and MgO is about 0.3\,\% and MgO is apparently well suited as substrate for both.\\
The AF/F exchange coupling was first reported in 1957 in Co/CoO systems\cite{FirstExchangeCoupling}. This pinning effect increases the coercive field of the F film and leads to a shift of the hysteresis loop by a bias field, the so called exchange bias. The pinning effect can be induced by cooling the AF/F bilayer from high temp through the N$\rm \acute{e}$el temperature of the AF under the application of a magnetic field.\\
The exchange bias on Fe$_{3}$O$_{4}$ films has been investigated in prior studies, i.e. by using different substrates to change the growth direction of the AF and analyzing the influence of a compensated NiO(001) and a fully uncompensated NiO(111) interface on the exchange bias\cite{Fertbilayer}. The influence of the stacking order was also examined in that study.\\
Models have been developed to describe the exchange bias, i.e a domain state model (DSM) based on Monte Carlo simulations\cite{DomainI}, which shows that the magnitude of the exchange bias is dependent on the degree of dilution of the antiferromagnet. This domain state model has been verified by experiments\cite{DomainII}.\\
Shortly after the exchange coupling, the trainings effect was found,too. The trainings effect denotes the descending of the initial exchange bias to a smaller residual value during measuring the hysteresis loops. D. Paccard has reported about the training effect\cite{TrainingsEffect1} in AF/F systems.\\
Since the film thickness, the crystal quality of the films and AF/F interface structure and roughness strongly influence the magnetic properties of the bilayers, we have investigated the thickness dependence of the strutural quality of Fe$_{3}$O$_{4}$/NiO-bilayers on the film thickness of each film. Both NiO and Fe$_{3}$O$_{4}$ were grown at 250$^{\circ}$ substrate temperature to avoid the unwanted interdiffusion of magnesium ions from the substrate into the bilayer\cite{PressureRate}. A substrate temperature of 250$^{\circ}$ is also considered as the lower limit for the growth of well-ordered magnetite films\cite{NiOMagnetit}.\\ 
The stoichiometry of the bilayers were analyzed $in-situ$ by x-ray photoelectron spectroscopy (XPS), while the structural properties were investigated by $in-situ$ low energy electron diffraction (LEED) and $ex-situ$ x-ray diffraction (XRD). The XRD data were evaluated using kinematic diffraction theory.\\

\section{EXPERIMENTAL SETUP AND SAMPLE PREPARATION}
The sample preparation was carried out in multi chamber ultra high vacuum (UHV) systems with a preparation chamber (base pressure of 10$^{-8}$\,mbar) and an analysis chamber (base pressure of 10$^{-10}$\,mbar). The available $in-situ$ characterization methods are low energy electron diffraction (LEED) and x-ray photoelectron spectroscopy (XPS). The XPS system operates with an hemispherical analyzer and an Al K$_{\alpha}$ x-ray anode (1486.6 eV).
To perform reactive molecular beam epitaxy (RMBE) the preparation chamber is equipped with electron beam evaporators for nickel and iron, a heatable manipulator and an oxygen source.\\  
Before film deposition the MgO substrates were annealed at 400$^\circ$C in a 10$^{-4}$\,mbar oxygen atmosphere to obtain well-ordered and clean surfaces as proved by LEED and XPS, respectively.
After the preparation of the substrates NiO films were deposited using RMBE at 250$^\circ$C in a 10$^{-5}$\,mbar oxygen atmosphere. Afterwards, Fe$_{3}$O$_{4}$ films with constant thickness were grown on the NiO films also via RMBE at 250$^\circ$C in a 5$\times$10$^{-6}$ mbar oxygen atmosphere.
The thicknesses of the different films were in situ controlled by a quartz crystal microbalance.
Later ex situ XRR measurements were performed to prove the thickness of all films of the samples. 
Two different series of Fe$_{3}$O$_{4}$/NiO bilayers were grown on MgO(001) supports. In series one the NiO film thickness was modified, while the magnetite film thickness was kept constant. The bilayers of series two has a constant NiO film thickness, while the magnetite thickness was varied.\\
After each annealing step and each film deposition the samples were transferred to the analysis chamber for $in-situ$ LEED and XPS measurements.\\
After the sample preparation the samples were exposed to ambient conditions for diverse $ex-situ$ experiments. The film thickness and structure of the films was analyzed by XRR and XRD, respectively.
These experiments were carried out at Deutsches Elektronen-Synchroton (DESY), Hamburg at PETRA III beamline P08. This is an undulator beamline with a high heat-load double-crystal monochromator and large-offset monochromator to separate the beamline from the adjacent beamline. At the endstation a Kohzu 4S+2D type diffractometer is installed\cite{OSeeckP08}. A Mythen array detector\cite{Detektor} was used due to its higher dynamic range and the capability of creating reciprocal space maps (RSM) within a shorter period of time compared to a point detector.

\section{Results}

\subsection{$In-Situ$ surface characterization by LEED and XPS}
\begin{figure}
	\centering
		\includegraphics{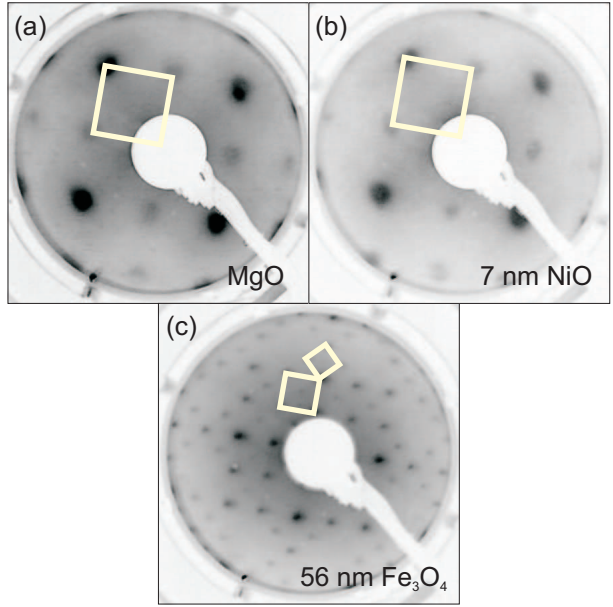}
	\caption{LEED pattern obtained at 135\,eV electron energy of a clean MgO(001) substrate (a), a NiO film on top of this substrate (b) and an Fe$_{3}$O$_{4}$ film on top of this NiO film. The small white square indicate the ($\sqrt{2} \times \sqrt{2})$R45$^{\circ}$ surface superstructure of the magnetite, while the bigger white squares indicate the 1 $\times$ 1 surface structures.} 
	\label{LEED}
\end{figure}

LEED experiments were performed at an electron energy of 135\,eV. Fig. \ref{LEED} shows exemplary for one sample the surface structure of an initial clean MgO substrate (a), with an additional 7\,nm NiO film (b) and the subsequently grown 56\,nm Fe$_{3}$O$_{4}$ film (c).
All bilayers show in each deposition step the same diffraction pattern.
Since MgO and NiO have the same rocksalt structure and similar lattice constants, the surface diffraction pattern is in either case a quadratic 1 $\times$ 1 structure.
The diffraction pattern of the magnetite film shows more spots due to the almost doubled lattice constant.
In addition, we always observe the typical ($\sqrt{2} \times \sqrt{2})$R45$^{\circ}$ superstructure.
This superstructure is only visible in the case of magnetite\cite{Cham1}, while maghemite has only an quadratic 1 $\times$ 1 surface structure.

\begin{figure}
	\centering
		\includegraphics{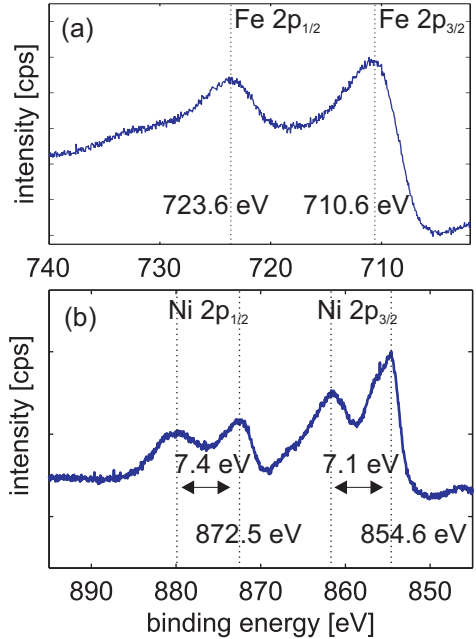}
	\caption{XP spectra of the Fe 2p (a) and the Ni 2p (b) of a 50\,nm iron oxide and a 7\,nm nickel oxide film, respectively.}
	\label{XPSCombo}
\end{figure}

The XP spectra of the Fe 2p peak region (a) and the Ni 2p region (b) are presented exemplarily in Fig.\ref{XPSCombo} for the same sample. 
The Ni 2p spectra consists of the Ni 2p$_{3/2}$ peak at 854.6\,eV and the Ni 2p$_{1/2}$ peak at 872.6\,eV with their corresponding satellites at about 7\,eV higher binding energy. The measured Ni 2p spectra agree well with Ni 2p spectra for NiO reported in literature\cite{NiOXPS}. The XP spectra of the Fe 2p region of all samples exhibit the same behavior. The Fe 2p$_{3/2}$ is located at binding energy of 710.6\,eV, while the Fe 2p$_{1/2}$ is situated at binding energy of 723.6\,eV. Magnetite contains Fe$^{2+}$-ions as well as Fe$^{3+}$-ions at the ratio of 1:2. At this ratio the charge transfer satellites characteristic for maghemite (718.8\,eV\cite{Yamashita,Fuji}) or wuestite (714.7\,eV\cite{Yamashita,Fuji}) are not visible separately since both satellites overlap forming a flat plateau between the Fe 2p$_{3/2}$ and Fe 2p$_{1/2}$ peak. Hence, this spectrum is typical for stoichiometric Fe$_{3}$O$_{4}$. Therefore, combined with the LEED experiments it is safe that  crystalline and stoichiometric magnetite has been deposited on crystalline and stoichiometric NiO. 

\subsection{Film structure characterization by XRD}

The structures of the entire oxide bilayers were analyzed by X-ray diffraction experiments.
Hence, the 3D reciprocal space spanned by the MgO substrate is indexed lateral by the MgO(001) surface unit cell while the layer distance has been used for the direction perpendicular to the surface. Compared to the well-known cubic bulk unit cell this surface unit cell has half the size of the bulk unit cell of MgO in vertical direction due to the spacing between (001) crystal planes and is rotated by 45$^{\circ}$ in the (001) plane.
Thus, compared to the cubic bulk lattice, we use the base vectors $\vec{a}_{1}\,=\,\frac{1}{2}(1,1,0)$, $\vec{a}_{2}\,=\,\frac{1}{2}(-1,1,0)$ and $\vec{a}_{3}\,=\,\frac{1}{2}(0,0,1)$ to describe the lattice.
As a consequence the MgO(002)$_{B}$ rock salt bulk reflection is denoted by MgO(001)$_{S}$.
Since the magnetite and the maghemite spinel structures have almost doubled bulk lattice constants compared to MgO, the (004)$_{B}$ spinel bulk reflection is very close to the MgO(001)$_{S}$ reflection 
NiO crystallizes like MgO in the rock salt structure with a slightly smaller lattice constant, so the NiO(002)$_{B}$ rock salt bulk reflection is also close to the MgO(001)$_{S}$ reflection.\\

\begin{figure}
	\centering
		\includegraphics{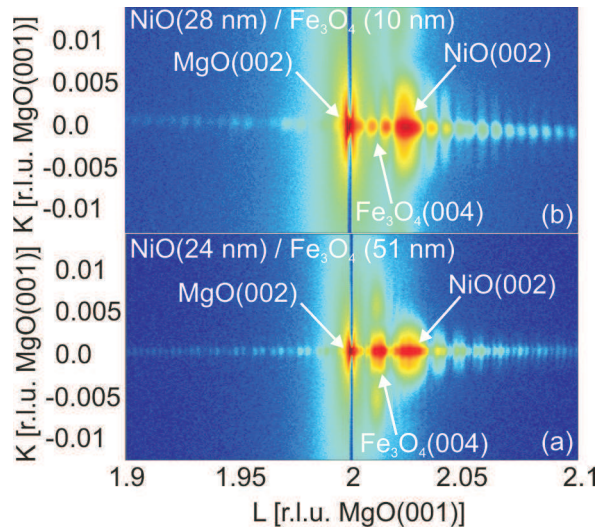}
	\caption{Reciprocal space map of a sample with a 24\,nm NiO and a 51\,nm Fe$_3$O$_4$ film (a) and of a sample with 29\,nm NiO and 10\,nm Fe$_3$O$_4$ film (b).}
	\label{SRSM}
\end{figure}

A reciprocal space ($K$,$L$) mapping (RSM) of one sample of the first series with 50\,nm Fe$_{3}$O$_{4}$ on 24\,nm NiO is shown in Fig.\ref{SRSM}(a). For comparison another RSM of a sample of the second series with a 28\,nm NiO film and a 10\,nm Fe$_{3}$O$_{4}$ film is depicted in Fig.\ref{SRSM}(b). Both RSMs show a sharp MgO substrate peak at $L$\,=\,2 due to diffraction at the MgO substrate. The sample of the first series in Fig.\ref{SRSM}(a) shows broad Bragg peaks and corresponding Laue oscillations of the deposited NiO and Fe$_{3}$O$_{4}$ films due to their finite film thickness. As opposed to this the sample of the second series in Fig.\ref{SRSM}(b) exhibits only the Bragg peak of the NiO film along with distinct Laue oscillations. Here, the diffraction signal of the iron oxide causes only weak modulations of the diffraction signal of the NiO since the Fe$_{3}$O$_{4}$ film is very thin.

\begin{figure}
	\centering
		\includegraphics{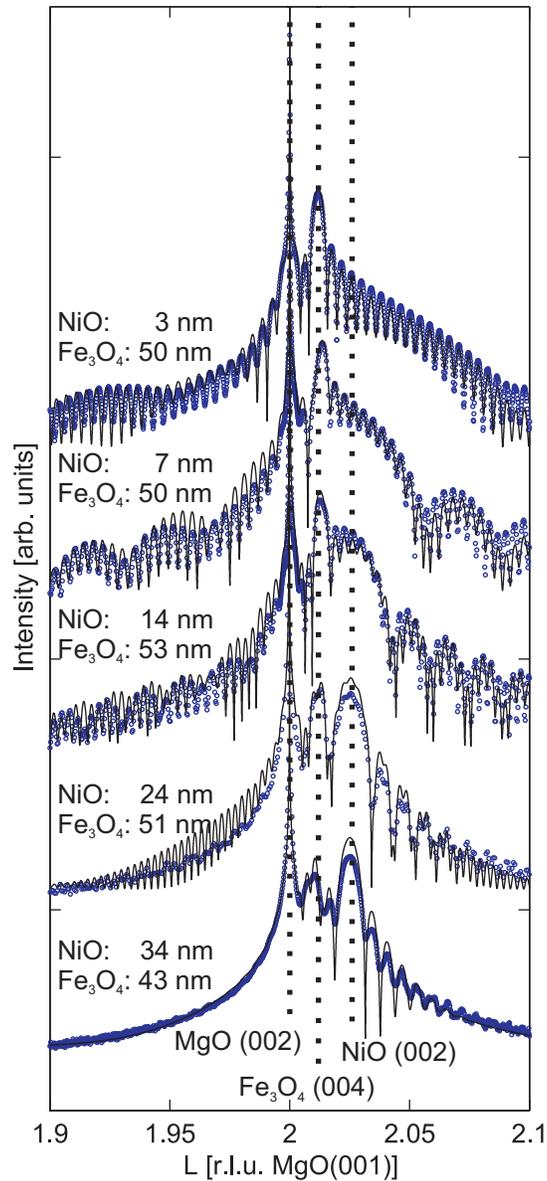}
	\caption{XRD rod scans along the (00$L$) of the samples with constant Fe$_{3}$O$_{4}$ thickness and increasing NiO thickness. Circles show experimental data and solid lines calculations.}
	\label{fig:XRD1}
\end{figure}

The crystal truncation rod (CTR) scans along the (00$L$) direction for the series with the nearly constant magnetite film and variation of the NiO film thickness are depicted in Fig. \ref{fig:XRD1}.
All scans show a sharp MgO substrate peak at $L$ = 2 due to diffraction at the MgO substrate. The series with the constant magnetite film thickness in Fig. \ref{fig:XRD1} exhibits in all scans broad Bragg peaks of the deposited Fe$_{3}$O$_{4}$ films due to their finite film thickness.
In addition well-pronounced Laue oscillations with high periodicity can be observed, which indicate well ordered films with homogeneous thickness.
The Bragg peaks of NiO are also visible for NiO thicknesses higher than 7 \,nm. The increasing film thickness of the NiO films can be easily seen by comparing the periodicity of the oscillations and the FWHM of the corresponding Bragg peaks (Fig. \ref{fig:XRD1}). However, at film thicknesses smaller than 7\,nm no Bragg peaks and only Laue oscillations of the NiO modulating the diffraction signal of the magnetite can be observed. At a NiO film thickness of 24\,nm the Laue oscillations of the magnetite film becomes weaker and are nearly vanished at a NiO film thickness of 33 nm. This observation indicates an increasing roughness of the NiO/Fe$_{3}$O$_{4}$-interface.
Moreover, here the Bragg peak of the magnetite film is weaker than the Bragg of the NiO film at 33\,nm although the  Fe$_{3}$O$_{4}$ film has a film thickness of 43\,nm.\\

\begin{figure}
	\centering
		\includegraphics{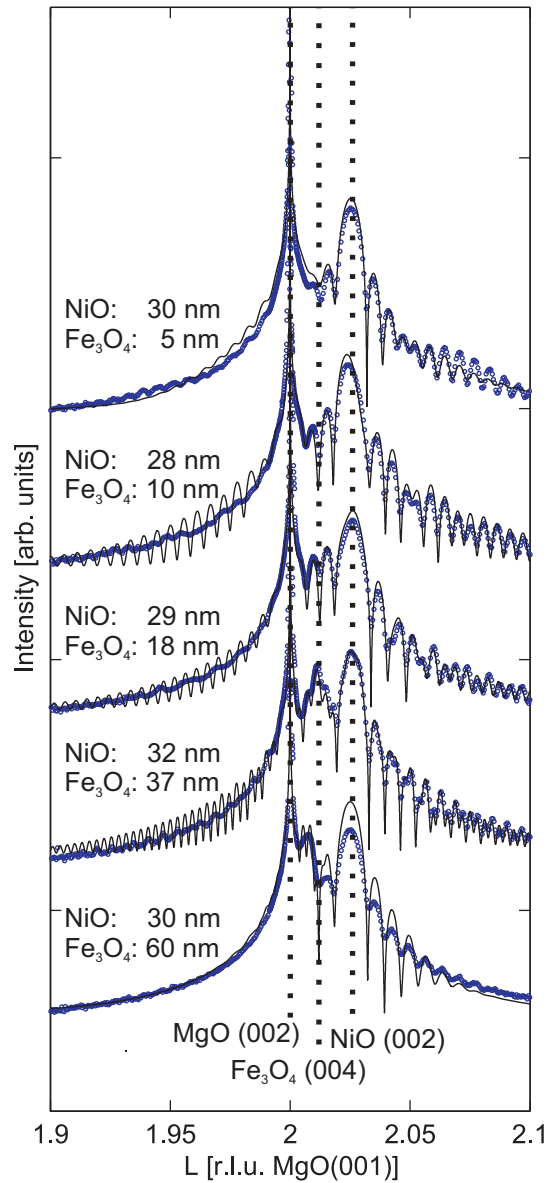}
	\caption{XRD rod scans along the (00$L$) of the samples with constant NiO thickness and increasing Fe$_{3}$O$_{4}$ thickness. Circles show experimental data and solid lines calculations.}
	\label{fig:XRD2}
\end{figure}

The specular rod scans of the second series of bilayers with a constant NiO film thickness of approximately 30\,nm are shown in Fig. \ref{fig:XRD2}. All films feature Laue fringes and broad Bragg peaks of the NiO film independent of the Fe$_{3}$O$_{4}$ film thickness.  In this series the Laue fringes for thin magnetite films cause very small modulations of the diffraction signal of the NiO. However, the thin magnetite films in this series modulate the diffraction signal of the NiO film much weaker than the thin NiO films modulate the thick magnetite films in series one. Another feature is that the Bragg peaks and fringes of the thicker magnetite films can hardly be seen. Even the Bragg peak of a 60\,nm thick magnetite film is almost not visible. One exception is the sample with a 37\,nm magnetite film, where some weak Laue fringes and a small Bragg peak are observable. However, overall the diffraction signal of the magnetite film is very weak compared to a well-ordered magnetite film. This observation indicates that the quality of the magnetite depends on the NiO film thickness. While small NiO thicknesses barely influence the crystal quality of the magnetite films, a NiO film thickness of 24\,nm increases the roughness of the NiO/Fe$_{3}$O$_{4}$-interface .\\

\begin{figure}
	\centering
		\includegraphics{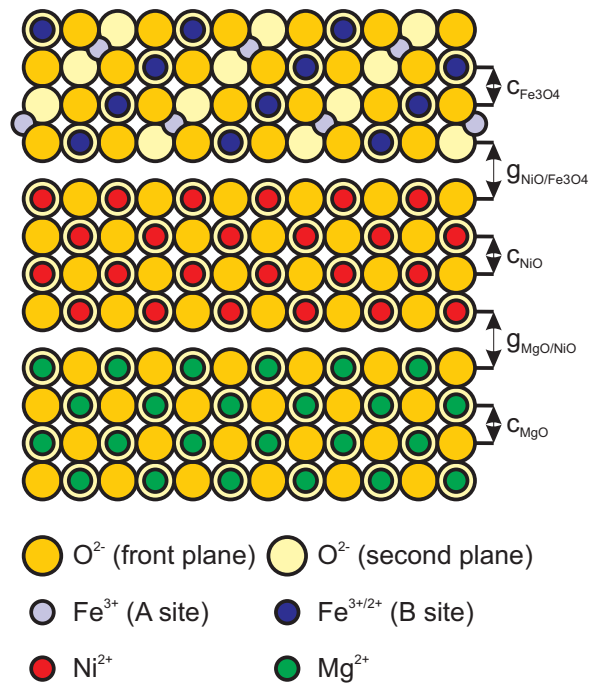}
	\caption{Schematic cross section through a sample to explain the parameters used in the layer model. The different colored circles represent the different ions in the corresponding crystal lattice.}
	\label{ModellNiFeO}
\end{figure} 

\begin{figure}
	\centering
		\includegraphics{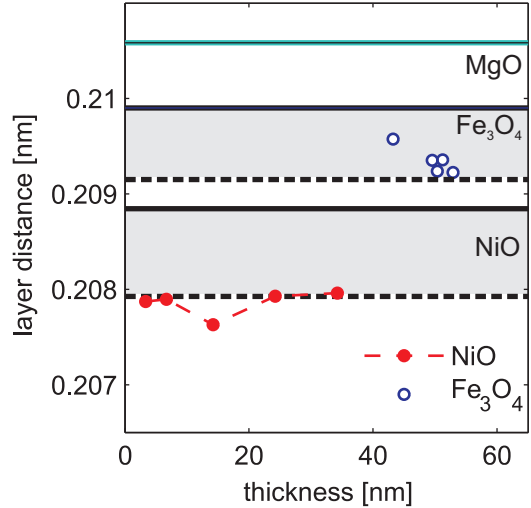}
	\caption{Vertical layer distance of magnetite and nickel oxide of the samples with constant Fe$_{3}$O$_{4}$ film thickness and increasing NiO film thickness dependent on film thickness.}
	\label{fig:Lattieserie1}
\end{figure}

\begin{figure}
	\centering
		\includegraphics{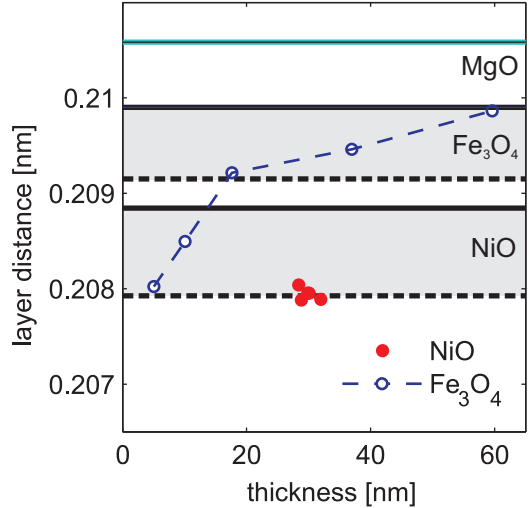}
	\caption{Vertical layer distance of magnetite and nickel oxide of the samples with constant NiO film thickness and increasing Fe$_{3}$O$_{4}$ film thickness dependent on film thickness.}
	\label{fig:Lattieserie2}
\end{figure}

We have applied full kinematic diffraction theory for analysis of the diffraction data lines in Figs. \ref{fig:XRD1} and \ref{fig:XRD2} in order to determine the structural parameters and to understand deeper the observations described above.
The applied approach for the calculations shown in Fig.\ref{ModellNiFeO} consists of the MgO rock salt substrate, the on top grown NiO with rock salt structure and the subsequently deposited Fe$_{3}$O$_{4}$ with spinel structure.
In this approach oxygen, Ni and iron ions were primarily arranged in their respective bulk structures and the diffracted intensity was calculated using their atomic form factors.
In the calculation the unit cells of the respective films were homogeneously deformed perpendicular to the surface to obtain the vertical layer distance.
Further parameters of the calculation are the surface roughness of the Fe$_{3}$O$_{4}$ film and interface roughnesses as well as the Debye-Waller factors.
The vertical layer distances determined from the curve fitting calculations of the (00$L$)-rod plotted against the film thickness are shown in Fig.\ref{fig:Lattieserie1} and Fig.\ref{fig:Lattieserie2}.

The vertical layer distances $c$ for completely strained pseudomorphic Fe$_{3}$O$_{4}$ and NiO were calculated using $\frac{\Delta c}{c} = \frac{2\nu}{\nu - 1} \frac{\Delta a}{a}$\cite{NiOPoisson} and assuming a Poisson number of $\nu$\,=\,0.356 for magnetite\cite{Poisson} and $\nu$\,=\,0.21 for NiO\cite{NiOPoisson}.
The areas between these calculated values (dotted lines) and their corresponding bulks values (solid lines) are marked in grey.
The vertical layer distances of NiO ($c_{1,NiO}$ ) and Fe$_{3}$O$_{4}$ ($c_{1,Fe_{3}O_{4}}$) for the series with increasing NiO film thickness are shown in Fig.\ref{fig:Lattieserie1}.
The vertical layer distance $c_{1,NiO}$ of the NiO (red dots) is completely strained (0.2079\,nm) and exhibits no dependence on the NiO film thickness.
Therefore, the NiO does not relax with increasing film thickness in the investigated range of film thicknesses. The vertical layer distances of magnetite is also strained and nearly constant at 0.2098\,nm and is consequently also independent of the NiO film thickness.

A similar observation can be made in Fig.\ref{fig:Lattieserie2}, where the vertical layer distances $c_{2,NiO}$ and $c_{2,Fe_{3}O_{4}}$ of the second series are plotted against the film thickness. 
In agreement with the first series of bilayers the vertical layer distances $c_{2,NiO}$ of the NiO films with constant film thickness are completely strained at approximately 0.2079\,nm.
The vertical layer distance $c_{2,Fe_{3}O_{4}}$ of Fe$_{3}$O$_{4}$ is heavily strained and relaxes distinctly from 0.2080\,nm at 5\,nm to 0.2092\,nm at 18\,nm and then slowlier to 0.2099\,nm for 60\,nm Fe$_{3}$O$_{4}$ film thickness.
Thus, magnetite reaches the vertical layer distance of bulk magnetite.
However, we have to admit that the vertical layer distance of the magnetite films in this series were more difficult to determine than in the other series since the intensity of the corresponding Bragg peaks and fringes were quite weak. The weak Fe$_{3}$O$_{4}$ Bragg peaks in this series confirms the observation for the first series that the quality of the magnetite films depends strongly on the NiO film thickness. Beyond 24\,nm NiO film thickness the magnetite film is not well-ordered.

\begin{figure}
	\centering
		\includegraphics{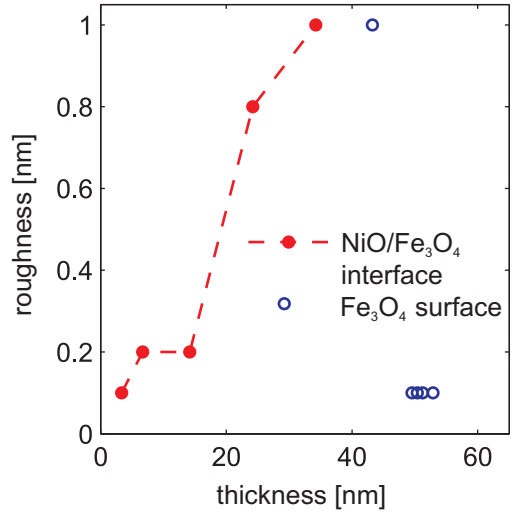}
		\caption{Surface and interface roughness of the samples with constant Fe$_{3}$O$_{4}$ film thickness and increasing NiO film thickness dependent on film thickness. }
	\label{fig:FeNiOS1roughness}
\end{figure}

This finding is strongly supported by our results for the NiO/Fe$_{3}$O$_{4}$ interface roughness. First, we have to emphasize tha the MgO/NiO interface is very smooth. It's roughness is 0.2($\pm$ 0.1)\,nm 
The roughness of the NiO/Fe$_{3}$O$_{4}$ interface and the Fe$_{3}$O$_{4}$ surface is plotted against film thickness in Fig.\ref{fig:FeNiOS1roughness}.
It can be seen that with increasing NiO film thickness the roughness of the interface is also continuously increasing from 0.1\,nm to 1\,nm. The surface of the magnetite films, however, exhibits a small roughness of 0.1\,nm independent of the film thickness with the exception of the Fe$_{3}$O$_{4}$ film with 33\,nm NiO underneath. Here, the magnetite film has also a roughness of 1\,nm. 

\begin{figure}
	\centering
		\includegraphics{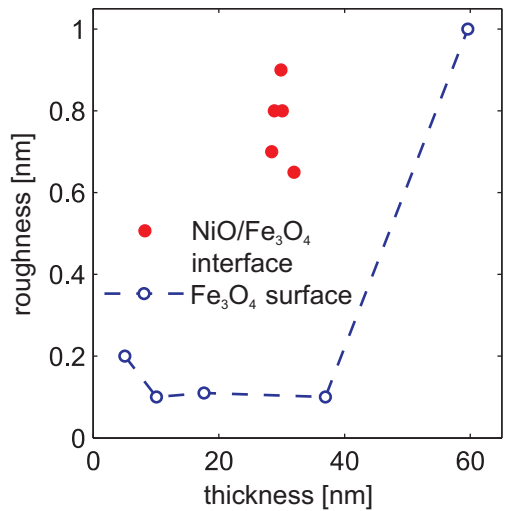}
		\caption{Surface and interface roughness of the samples with constant NiO film thickness and increasing Fe$_{3}$O$_{4}$ film thickness dependent on film thickness. }
	\label{fig:FeNiOS2roughness}
\end{figure}

\begin{figure}
	\centering
		\includegraphics{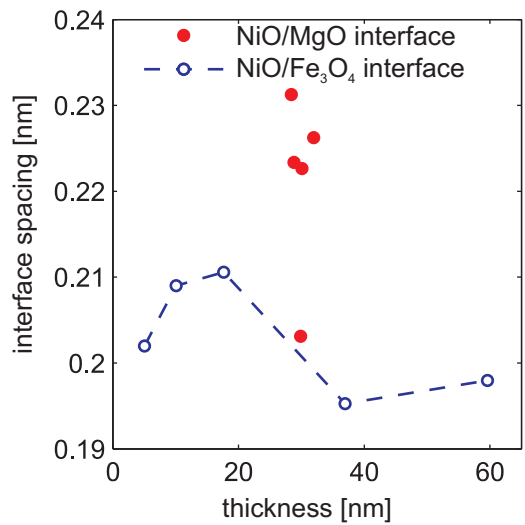}
		\caption{Interface spacing of the samples with constant NiO film thickness and increasing Fe$_{3}$O$_{4}$ film thickness dependent on film thickness. }
	\label{fig:FeNiOS2interface}
\end{figure}

The interface and surface roughness of the NiO and Fe$_{3}$O$_{4}$ films of the second series of samples are shown in Fig. \ref{fig:FeNiOS2interface}. In agreement with the first series the roughness of the NiO/Fe$_{3}$O$_{4}$ interface  is 0.80 ($\pm$ 0.15)\,nm for the 30$\pm 2$\,nm NiO films, while the surface roughness of the magnetite films is 0.15($\pm$0.05)\,nm with the exception of the magnetite film with 60\,nm film thickness. Here, interface roughness is 1\,nm.   

\begin{figure}
	\centering
		\includegraphics{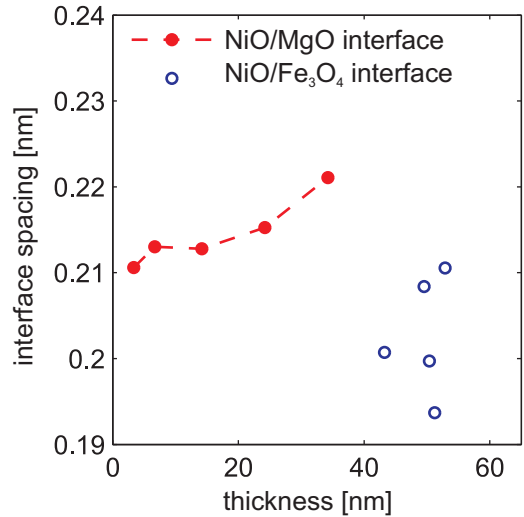}
		\caption{Interface spacing of the samples with constant Fe$_{3}$O$_{4}$ film thickness and increasing NiO film thickness dependent on film thickness. }
	\label{fig:FeNiOS1interface}
\end{figure}

The spacings for the MgO/NiO and NiO/Fe$_{3}$O$_{4}$ interfaces (cf. Fig.\ref{ModellNiFeO}) are depicted dependent on film thickness in Fig. \ref{fig:FeNiOS1interface} for the first series.
The interface spacing MgO/NiO is slighty increasing with increasing film thickness from 0.2106\,nm to 0.2211\,nm, while interface spacing NiO/Fe$_{3}$O$_{4}$ is 0.202($\pm$0.008)\,nm and shows no dependence on film thickness.

The spacings of the MgO/NiO and NiO/Fe$_{3}$O$_{4}$ interfaces of the second series of samples are plotted versus film thickness in Fig. \ref{fig:FeNiOS2interface}. Both spacings show no dependence on film thickness and are 0.206($\pm$ 0.005)\,nm for the NiO/Fe$_{3}$O$_{4}$ interface while the MgO/NiO interface spacing is 0.217($\pm$ 0.014)\,nm.   

\section{discussion}
After preparation of the nickel oxide and iron oxide films the photoelectron spectra of the Fe2p and Ni2p peak reveal that the surface near region of the films is Fe$_{3}$O$_{4}$ and NiO, respectively.
Both NiO and Fe$_{3}$O$_{4}$ films show LEED diffraction patterns with the expected surface structures. NiO films exhibit a (1 $\times$ 1) surface structure, since they crystallize like MgO in rock salt structure. Fe$_{3}$O$_{4}$ films have the the typical ($\sqrt{2}\times\sqrt{2}$)R45$^{\circ}$ superstructure of the surface.\\
The photoelectron spectra of the Fe2p and Ni2p peak reveal that the surface near region of the films is Fe$_{3}$O$_{4}$ and NiO, respectively. In summary, LEED and XPS prove that NiO and Fe$_{3}$O$_{4}$ films have the expected surface structure and surface near stoichiometry and there is no dependence on film thickness.\\
XRD specular rod scans were carried out to investigate the whole structure of the samples. The scans of all films reveal that all NiO films are crystalline and well-ordered.
In the first series where the initially NiO film thickness is 3\,nm only a weak broad Bragg peak can be observed, however, the strong Laue fringes of the NiO modulate the diffraction signal of the Fe$_{3}$O$_{4}$ film. A separate broad Bragg peak of the NiO film is observable at 14 \,nm and it gets more distinct with increasing NiO film thickness.
All NiO films in both series are fully strained with a vertical layer distance of 0.2079 nm and do not relax with increasing film thickness. The interlayer spacing of NiO normal to the surface has become smaller to compensate the tensile strain due to the lattice matching of the film with the surface unit cell of the MgO substrate. This strain should decrease rapidly with the stable formation of dislocation above the critical thickness $h_{c}$. The critical thickness $h_{c}$ for the formation of misfit dislocations can be calculated using the formula\cite{FormelcritThick}
\begin{align}
 \frac{h_{c}}{b}= \frac{1 - \nu \cdot \cos^{2}(\alpha) (ln(\frac{h_{c}}{b}) +1)}{2\pi f(1 + \nu) \cos(\lambda)} \,
\label{Gleichung} ,
\end{align}
where $b=\frac{a_{NiO}}{\sqrt{2}}$ is the magnitude of the Burgers vector, $f$ = 0.8\% is the misfit of NiO, $\alpha = 90^{\circ}$ is the angle between the dislocation line and the Burgers vector, $\lambda = 45^{\circ}$ is the angle between the Burgers vector and the direction that is both normal to the dislocation line and that lies within the plane of the interface and $\nu = 0.21$ is the Poisson ratio\cite{NiOPoisson}. We obtain a critical film thickness of $h_{c}$ = 39\,nm, which means that the NiO film grown for this study are all below this critical film thickness and it is reasonable to observe no strain relaxation at the vertical layer distance. In addition, we like to emphasize that the relaxation process is very slow also for reasonable thicker NiO films\cite{NiOPoisson}.\\
The determined interface spacings $g$ between the MgO substrate and the NiO film and between the NiO and the Fe$_{3}$O$_{4}$ (cf. Fig. \ref{ModellNiFeO}) show no interpretable dependence on the film thickness. Only the spacing of the MgO/NiO interface of the first film series increases slightly with increasing film thickness. While the Fe$_{3}$O$_{4}$/NiO interface thickness does not differ significantly from the interlayer spacings of these oxide films the NiO/MgO interface thickness is expanded compared to their interlayer spacings. This may be attributed to some covalent character of NiO - MgO interactions\cite{NiOMgOcovalent}. \\
The roughness of the NiO/Fe$_{3}$O$_{4}$ interface in series one is increasing with increasing NiO film thickness from 0.1\,nm to 1\,nm. In series two the NiO/Fe$_{3}$O$_{4}$ interface roughness is nearly constant at 0.80 ($\pm$ 0.15)\,nm. This is consistent with series one since the constant NiO film thickness in series two confirms approximately the NiO film thickness of the thickest NiO film in series one.\\
Gatel et al. have grown among others Fe$_{3}$O$_{4}$/NiO bilayers on MgO(001) with a constant NiO film thickness of 66\,nm and different Fe$_{3}$O$_{4}$ film thicknesses ranging from 5 - 100\,nm. NiO and magnetite films have been grown using RF-sputtering at 700$^{\circ}$C and at 400 $^{\circ}$C substrate temperature, respectively. Gatel et al. state a NiO/Fe$_{3}$O$_{4}$ interface roughness of 0.7\,nm of their bilayers on MgO(001)\cite{Fertbilayer}.
Thus, the obtained interface roughness is in good agreement with literature.
Growth studies of single NiO films on MgO(001) at different preparation temperatures ranging vom 500$^{\circ}$C to 900$^{\circ}$C have shown that the surface roughness of NiO films becomes higher with increasing growth temperature from 0.2\,nm to 5.0\,nm \cite{CoNiObilayer}. In both studies the NiO films were grown using sputter deposition. Thus, lower growth temperatures reduce the interface roughness. In another study James et al. have grown NiO films with different thickness (20\,nm - 162\,nm) on MgO(001) using NO$_{2}$ assisted RMBE. They obtain an average NiO surface roughness of 0.35\,nm. 
The latter study confirms that our growth temperature should not be the reason for the rough NiO/Fe$_{3}$O$_{4}$ interface, since a growth temperature of 250$^{\circ}$ is sufficient to obtain smooth surfaces even at 162\,nm NiO film thickness.
One possible explanation is that the interface roughness is increasing during the deposition of magnetite leading to a high NiO/Fe$_{3}$O$_{4}$ interface roughness since intermixing effects may play an important role at least at higher temperatures. However, regarding the development of the roughness in the first series, this should not be the case since the interface roughness is increasing with growing NiO film thickness.\\  
 While the structural quality of the NiO films is constantly high in both series, the structural quality of the Fe$_{3}$O$_{4}$ film gets worse with increasing NiO film thickness. Until 24\,nm NiO film thickness the Fe$_{3}$O$_{4}$ exhibits an obvious Bragg peak with corresponding Laue fringes.
Above this thickness only weak Fe$_{3}$O$_{4}$ Bragg peaks are visible and nearly no Laue fringes can be seen. This observation at the first series is confirmed by CTR analysis of the second series, where the film thickness of all NiO films is approximately 30\,nm. While at small Fe$_{3}$O$_{4}$ film thickness the Laue fringes of magnetite causes weak modulations of the NiO diffraction, no distinct Bragg peak corresponding to Fe$_{3}$O$_{4}$ appears between the NiO and MgO Bragg peaks with increasing Fe$_{3}$O$_{4}$ film thickness. The Laue fringes corresponding to magnetite do not reach a comparable strength to the Laue fringes of the first three samples of the first series. So we have to note that the NiO film thickness has an influence on the structural quality of the magnetite films. The reason for the bad structural ordering of the magnetite is obviously the high NiO/Fe$_{3}$O$_{4}$ interface roughness which is increasing with advancing NiO film thickness (Fig. \ref{fig:FeNiOS1roughness}). 
Calculating the critical film thickness h$_{C}$ for the formation of misfit dislocations in magnetite using formula \ref{Gleichung} we obtain h$_{C}$ = 105\,nm ($b = 0.2969\,nm$, $f = 0.3\,\%$, $\nu = 0.356$, $\lambda = 45^{\circ}$, $\alpha = 90^{\circ}$ ), which is approximately twice the film thickness of magnetite in this study. Since NiO grow pseudomorph on MgO adapting its lateral lattice constant, the misfit $f$ is respective to the MgO. \\ 
In the first series of bilayers the vertical layer distance of magnetite features no dependence on NiO film thickness, which is reasonable since the magnetite film thickness of every bilayer is almost the same and the NiO films are pseudomorph to MgO. However, in the second series the vertical layer distance of Fe$_{3}$O$_{4}$ relaxes with increasing Fe$_{3}$O$_{4}$ film thickness until it reaches nearly the bulk value, although the magnetite have not reached the critical film thickness. Our previous studies on the growth of magnetite on MgO(001)\cite{BertiBernd,Berti2heizen,Berti1} confirm this observation. In these studies the strain of the magnetite films has relaxed with increasing film thickness and the critical film thickness is very small.\\
Although the structural ordering of the magnetite gets worse, the roughness of the magnetite surface is relative low ranging from 0.1\,nm to 0.2\,nm. Thus, magnetite films seem to compensate the interface roughness. Only the bilayer with 33\,nm NiO film thickness in series one and the bilayer with 60\,nm magnetite film thickness in series two have a distinct higher roughness with 1\,nm. This can be attributed to the great progression of the structural degradation of the magnetite films. Thus, in order to get structural better magnetite films on thick NiO films, the preparation of the Fe$_{3}$O$_{4}$/NiO bilayers has to be improved. Therefore, the influence of the growth temperatures of NiO and Fe$_{3}$O$_{4}$ on the Fe$_{3}$O$_{4}$-NiO interface roughness have to be investigated. As mentiond above a growth temperature of 500$^{\circ}$C leads to a smooth NiO film surface. Withal you have to keep in mind that a too high growth temperature may lead to an intermixing of the NiO and Fe$_{3}$O$_{4}$ films, which will also affect the magnetic properties of the bilayers.





\section{conclusion}
The detailed structural characterization of Fe$_{3}$O$_{4}$/NiO bilayers grown on MgO(001) substrates have shown that the quality of the NiO/Fe$_{3}$O$_{4}$ interface has a huge impact on the quality of the Fe$_{3}$O$_{4}$ films. While magnetite grows homogenously and smoothly on NiO films with up to 24\,nm thickness, the structural quality of the magnetite films gets distinctly worse with higher NiO film thickness. We attribute this to the fact that the interface roughness between NiO and Fe$_{3}$O$_{4}$ is increasing with increasing NiO film thickness. While the roughness of the 3\,nm NiO film is rather small, it is rising obviously with increasing NiO film thickness. As a result the structural quality of the magnetite films grown on 30\,nm NiO films is constantly reduced with increasing magnetite film thickness. \\

\section{acknowledgements}
Parts of this research were carried out at the light source PETRA III at DESY, a member of the Helmholtz Association (HGF).
We would like to thank O. H. Seeck for assistance in using beamline P08. Financial support by the DFG (KU2321/2-1) is gratefully acknowledged.

\end{document}